# Joint Sub-carrier and Power Allocation in MU-OFDM-DCSK Systems with Noise Reduction

M. Mobini, *M.R Zahabi*, [1] Electrical Engineering Department, Babol Noshirvani University of Technology, Iran

[*]zahabi@nit.ac.ir

***Abstract*—This paper investigates the joint number of sub-carriers and power optimization in Multi-user Orthogonal Frequency Division Multiplexing-based Differential Chaotic Shift Keying (MU-OFDM-DCSK) systems with noise reduction. We first find a closed-form expression for the optimal number of references in MU-OFDM-DCSK systems, while existing works only consider the numerical simulations. Moreover, this paper is the first to consider jointly sub-carrier and power allocation in the MU-OFDM-DCSK system by defining a fractional problem. The BER expression of the MU-OFDM-DCSK system as an objective function is non-convex and complex, which makes the analytical solution so difficult. We make it tractable by converting the fractional problem into a subtractive form, which leads to a near optimal solution. Simulation results show that our proposed approach provides better BER performance than the existing plans.**

***Index Terms*— MU-OFDM-DCSK, Joint Sub-carrier and power allocation, fractional programing.**

## I. Introduction

Many coherent and non-coherent chaos-based modulation schemes have been proposed so far [1-5]. The non-coherent chaotic modulations such as Differential Chaos Shift Keying (DCSK) [6] provide excellent performance, where a communication channel suffers from multi-path fading effect. More notably, the DCSK scheme can be implemented without any complex channel estimator and synchronizer [7]. In the DCSK modulation, a reference signal is transmitted followed by an information-bearing signal in two consecutive time slots. On the receiver side, these two signals are correlated whereby the information is obtained. A major weakness of the DCSK is that for half the bit duration it sends reference samples that carry no data. This can remarkably reduce its energy efficiency and data rate.

In order to improve the Bit Error Rate (BER) performance of the DCSK, Multi-Carrier DCSK (MC-DCSK) system has been suggested in [8]. On the transmitter side of the MC-DCSK, one transmitted symbol is spread over several orthogonal sub-carriers, and only one sub-carrier is used for transmitting the reference signal. In [9], the authors introduced the Orthogonal Frequency Division Multiplexing-based DCSK (OFDM-DCSK) system, in which the reference signals are arranged among the data-bearing sub-carriers. In order to improve the bandwidth efficiency and provide multi-user ability, the Multi-user OFDM-DCSK (MU-OFDM-DCSK) is proposed in [10]. In a recent study, authors showed that a non-coherent system can be more effective compared with the existing coherent candidates for 5G plans such as a millimeter-wave communication system, massive MISO, or similar circumstances that channel estimation is impractical [11]. This structure also is commodious for applications such as Wireless Sensor Network (WSN), Internet of Things (IoT), and ultra-wideband applications [12]. Although the MU-OFDM-DCSK offers higher performance in terms of energy efficiency and BER performance rather than the conventional DCSK, it is vulnerable to noise yet. Recently, Noise Reduction DCSK (NR-DCSK) system is designed, in which all the received reference signals are averaged to cope with the noise effects [13].

### A. Related Works

Several power allocation strategies have been proposed in multi-carrier-based DCSK systems, which are often based on the optimizing the BER expression [14-19]. Unfortunately, they are challenging non-convex and complex problems due to the nature of these systems [17-19]. In a recent paper [20], the sub-carrier allocation is considered and a sub-carrier-allocated MC-DCSK (SA-MCDCSK) system is proposed, in which several sub-carriers are adopted for reference signal transmission to obtain noise reduction through averaging. The authors formulated a BER minimization problem to find the optimal number of reference sub-carriers and confirmed it by numerical simulations. They do not produce a closed-form solution for the optimal number of reference sub-carriers.

Inspired by [20], we solved the sub-carrier allocation problem for an SA-OFDM-DCSK using Lagrangian relaxation to derive a closed-form solution. Moreover, we considered jointly optimization of the sub-carriers and their allocated power in the MU-OFDM-DCSK (PSA-OFDM-DCSK) that has not been addressed in the literature so far.

### B. Contributions

- As the first contribution, we derived an analytical solution for the optimal number of reference sub-carriers that minimizes the BER of MU-OFDM-DCSK system, while other articles (e.g. in [20]) only perform numerical simulations and did not present a closed-form expression for the optimal number of reference sub-carriers. Generally, the BER minimization problems of the multi-carrier-based DCSK systems have a high computational complexity, which makes the optimal solution intractable. For this reason, analytical solutions and closed-form formulas are not provided for the sub-carrier and power allocation in such systems [16-20]. We offered a methodology based on the Lagrangian relaxation to find a bound on a given problem. Sometimes, this bound is exactly matches the optimal solution [21].

- As the main contribution, we simultaneously considered the sub-carrier and power allocation in the MU-OFDM-DCSK systems, that has not been addressed in the literature so far. To this aim, we formulated a BER minimization problem as a single ratio fractional problem, and then, we offered a search-based method that converges to the global optimum solution.
- We also took into account the power of the reference sub-carriers, which are often relaxed in the existing studies. For example, in [17], [18], [19] the transmit power of reference sub-carrier has been assumed to be one for the problem simplification. In other words, the optimal power for reference sub-carrier has not been addressed in the literature so far.

The rest of this paper is organized as follows. Section II describes the proposed Sub-Carrier Allocated MU-OFDM-DCSK (SA-OFDM–DCSK) and derives a BER expression for it and obtains the optimal number of reference sub-carriers. In section III, the joint sub-carrier and power allocation policy and its related structure (PSA-OFDM–DCSK) is introduced. Simulation results are given in Section IV. Concluding remarks are presented in Section V.

## II. Sub-carrier-allocated MU-OFDM-DCSK (SA-OFDM-DCSK)

The BER optimization problem in the multi-carrier-based DCSK systems is challenging because the transmit power of the reference and data-bearing signals strongly impact each other through the noise terms in SNR. The SA-MC-DCSK system is introduced in [20], in which the authors focused on the optimal number of reference sub-carriers, and investigated it by numerical simulations. Inspired by [20], In this part, we first will describe the SA-OFDM-DCSK system model. Then, we will calculate the BER performance expression as our objective function. Finally, we will describe optimization policy to derive a closed-form solution for the number of reference subcarriers.

### A. System Model and signal format

In the MU-OFDM-DCSK system, the data sequence of the user $p^{th}$ is first formed into $M$ parallel sequences ($S_{1,P}$, $S_{2,p}$, …, $S_{i,p}$, …, $S_{M,p}$). A number of $N$ copies of the chaotic reference signal are transmitted while $M$ data bits are spread due to multiplication in the time with the chaotic spreading sequence $x_p(t)$:

$$x_p(t) = \sum_{k=1}^{\beta} x_{k.p}\, h(t - kM_c). \tag{1}$$

where β is the spreading factor, $h(t)$ is the shaping filter which is presumed to be rectangular and $M_c$ is the chip duration. In other words, the spreading process is done in time domain where $\beta$ number of IFFT operations are needed to prepare an SA-OFDM-DCSK symbol. Then, the parallel signal is transformed into a serial form and a cyclic prefix is added to it to reduce the Inter Symbol Interference (ISI) destructive effects. Therefore, the transmitted signal of the $p^{th}$ user in the SA-OFDM-DCSK system is given by:

$$e_p(t) = \sum_{v=1}^{N}\sum_{k=1}^{\beta} x_{k.P} e^{2\pi j f_{P_{pv}}(t-kM_c)} h(t-kM_c) + \sum_{\substack{i=1\\ i\neq p}}^{M}\sum_{k=1}^{\beta} x_{k.P} s_{i.P} e^{2\pi i f_{s_{pi}}(t-kM_c)} h(t-kM_c). \tag{2}$$

where $e_p(t)$ represents the transmitted OFDM-DCSK symbol of the user $p^{th}$, $f_{P_{pv}}$ is its $v^{th}$private frequency used to transmit the reference signal $x_{k.P}$, $f_{s_{pi}}$ is the $i^{th}$ shared frequency of the M frequencies to transmit the $i^{th}$ bit of the M block of bits. Fig. 1 denotes a transmitted frame of the SA-OFDM DCSK scheme which this format is similar for all users.

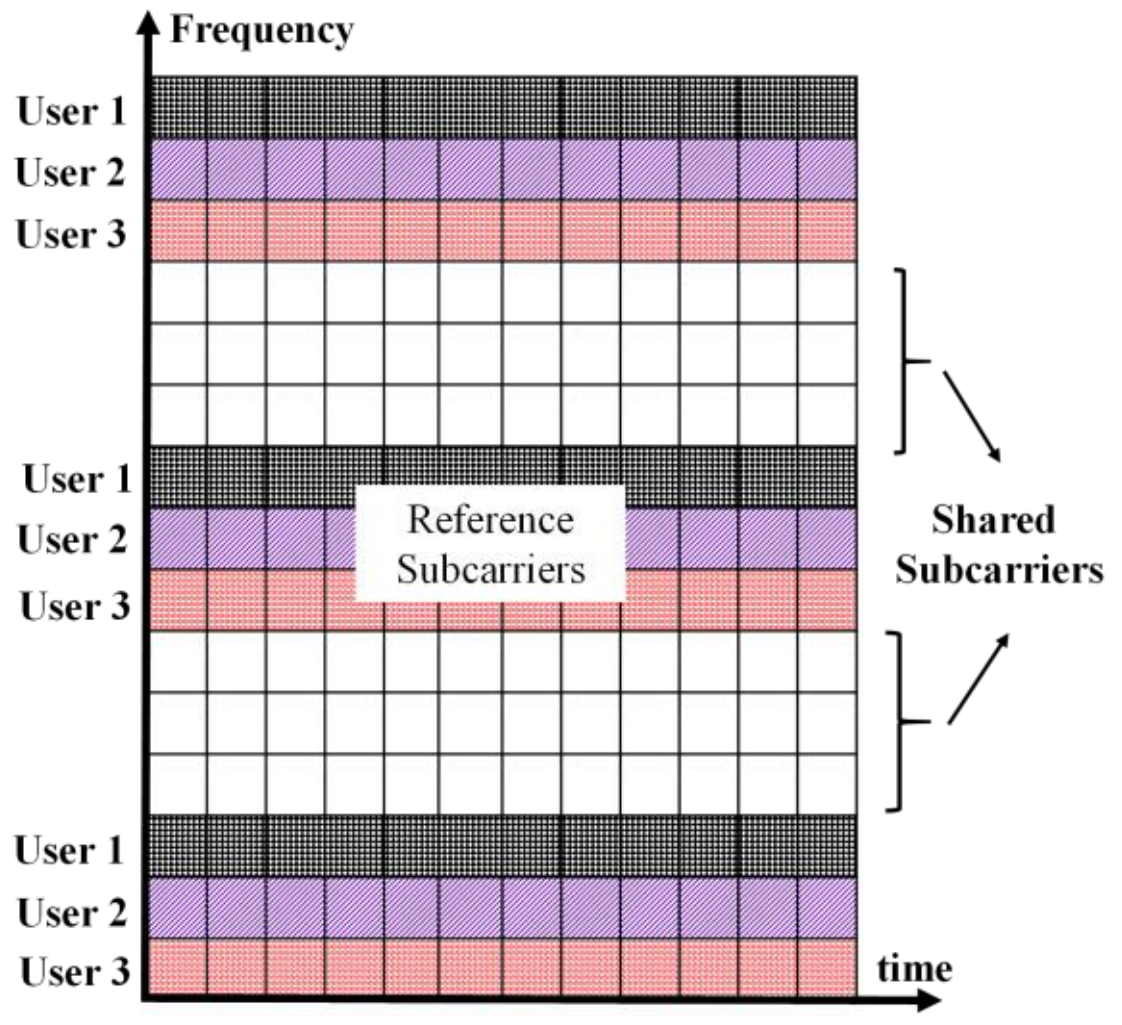

Fig. 1. Signal format for the p^th user.

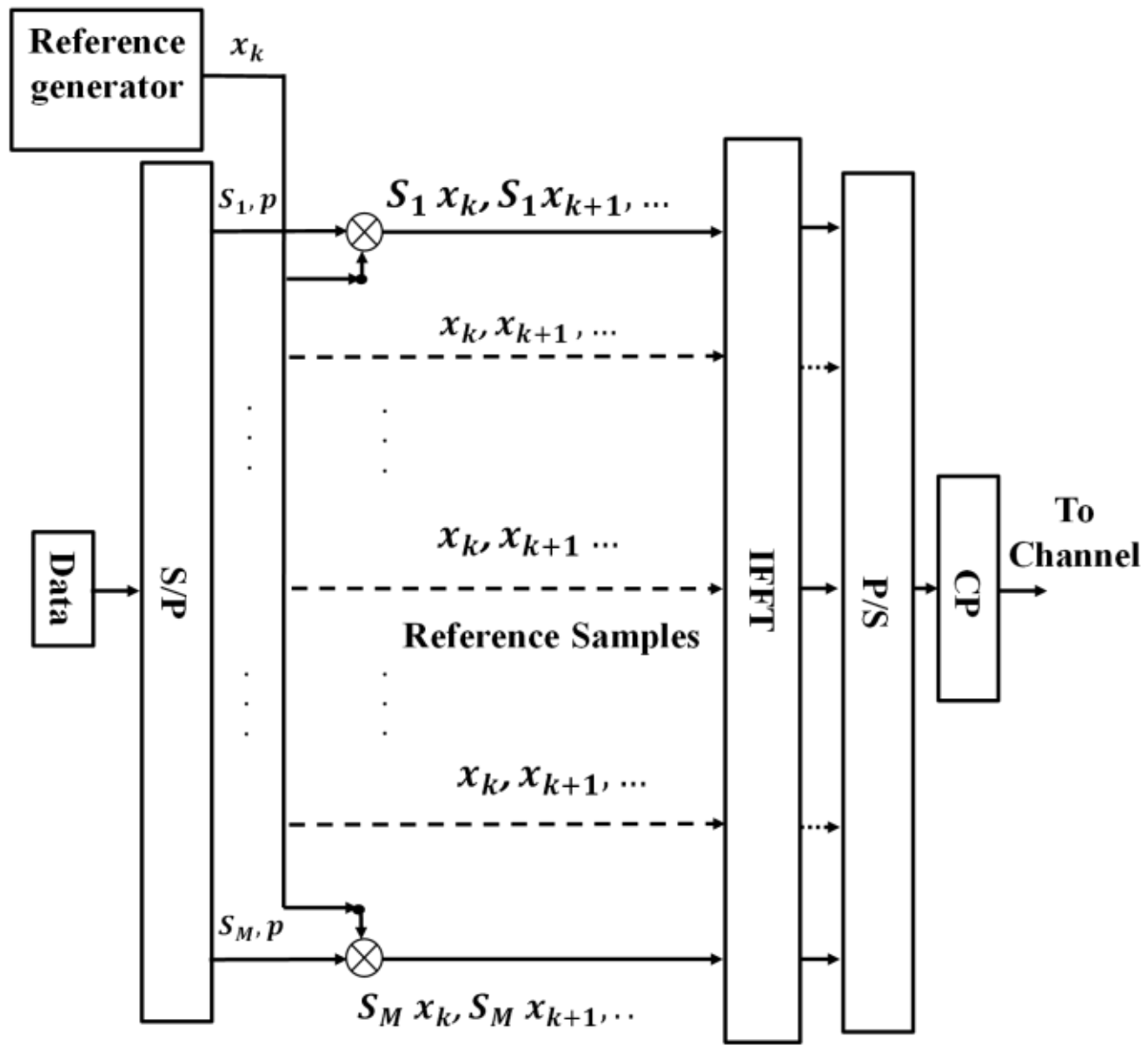

Fig. 2. Transmitter of the MU-OFDM-DCSK with noise reduction.

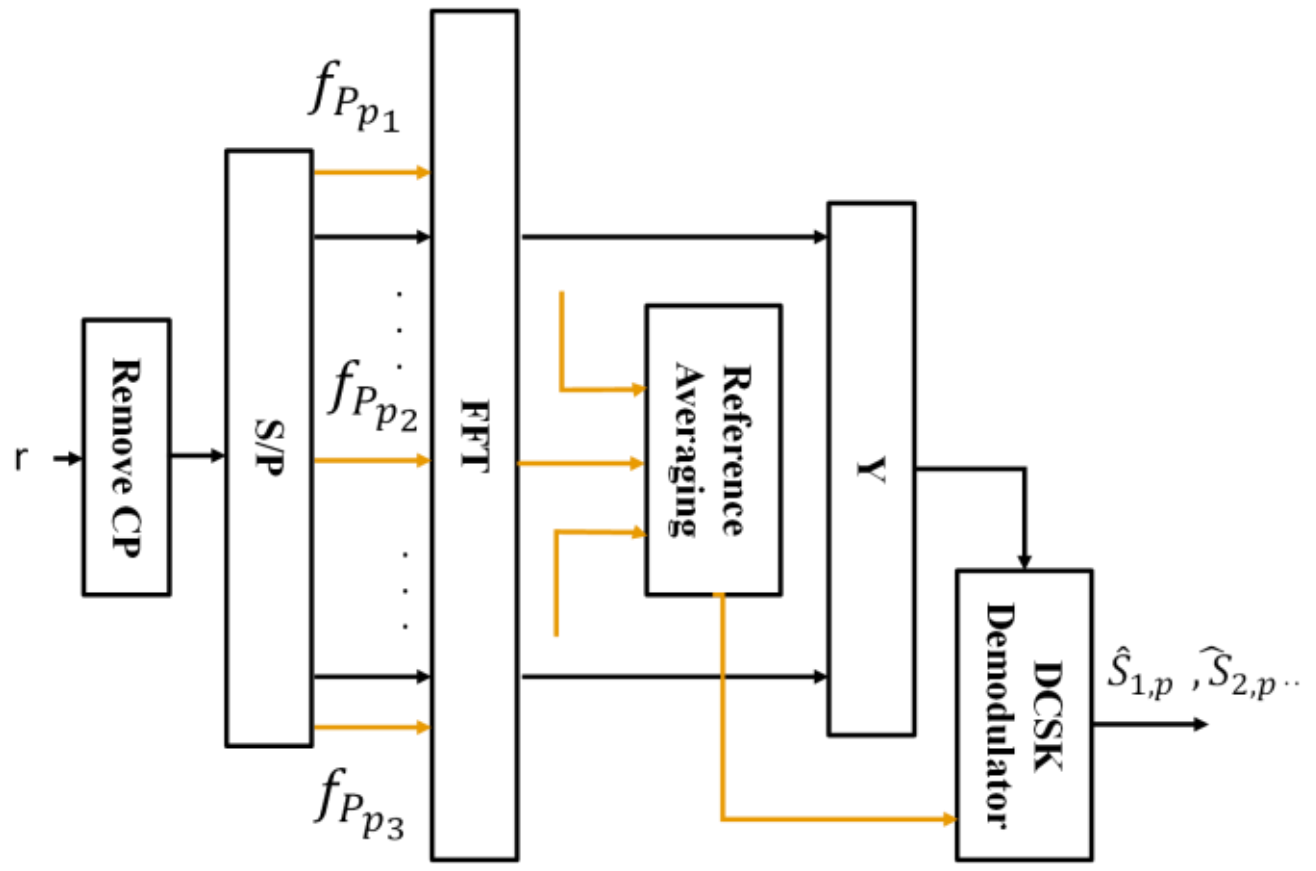

Fig. 3. Receiver of the MU-OFDM-DCSK with noise reduction.

As illustrated in Fig.2 and Fig.3, the general block diagram of the considered system is identical to that of the conventional MU-OFDM-DCSK, except for an averaging block at the receiver side. This unit is considered for each user and all received versions of the reference signal are averaged, and the averaged version will be used for the correct data detection.

### B. *BER analysis of the MU-OFDM-DCSK with noise reduction*

In this sub-section, we would like to obtain a BER expression for the MU-OFDM-DCSK with noise reduction as our optimization criterion. To this aim, we should first analyze the energy relations and define $E_b$ as the energy needed for the transmission of one data bit. As shown in Fig. 2, each $M$ bits require the number of $N$ of the chaotic reference signal where $N < M$. Hence, $N$ multiples of the reference energy $E_{ref}$ are required to transmit $M$ bits. Therefore, the total energy required to transmit one bit of the OFDM-DCSK with noise reduction system can be expressed as follows:

$$E_b = E_{data} + \frac{NE_{ref}}{M} = \left(\frac{M}{M-N}\right) M_c \sum_{k=1}^{\beta} x_k^2 \ . \quad (3)$$

Here, we get an equal energy for the reference and data sequences $E_{ref} = E_{data}$. If we assume that $M_c = 1$, (3) can be written as:

$$E_b = \left(\frac{M}{M-N}\right) \sum_{k=1}^{\beta} x_{k,p}^2 \ . \quad (4)$$

In this paper, we assume an AWGN channel. However, because the power allocated to the information-bearing sub-carriers were assumed to be equal, and only the number of the references will be optimized, the results could be extended to Rayleigh fading channel without loss of generality. This choice can help to better understand and summarize the formulas.

At the receiver side, since noise samples of different sub-carriers are independent and i.i.d., the variance of the noise at the output of the averaging block is $N$ times smaller than that of $n_{vk}$ (Gaussian noise added to the $k^{th}$ sample of the $v^{th}$ reference sub-carrier). To derive the BER expression of the user $p$, the mean and the variance for $i^{th}$ bit of the observation signal $D_{i.p}$ should be calculated. Therefore, the demodulator measures the following decision variable $D_{i.p}$:

$$D_{i.p} = Re\{\sum_{k=1}^{\beta} Y\left(k.i.f_{s_{pi}}\right).R(k)^*\}, \quad (5)$$

where $R$ is the averaged reference signal, and $Y\left(k.i.f_{s_{pi}}\right)$ is the spread data which are given by:

$$Y\left(k.i.f_{s_{pi}}\right) = \sum_{p=1}^{P} s_{i.p} x_{k.p} + n_{ik} \ , \quad (6)$$

$$R(k) = x_{k.p} + \frac{1}{N}\sum_{v=1}^{N} n_{vk} \ , \quad (7)$$

where $n_{ik}$ is Gaussian noise added to the $k^{th}$ sample of data-bearing signal placed in the $i^{th}$ sub-carrier. With replacing (6) and (7) in (5), and perform multiplication, the decision variable can be obtained as:

$$D_{i.p} = Re\{s_{i.p}\sum_{k=1}^{\beta} x_{k.p}^2 + \underbrace{\sum_{k=1}^{\beta} x_{p.k}\left(n_{ik} + s_{i.p}\left(\frac{1}{N}\sum_{v=1}^{N} n_{vk}\right)^*\right)}_{A} + \underbrace{\sum_{k=1}^{\beta}\left(\frac{1}{N}\sum_{v=1}^{N} n_{vk}\right)^* n_{ik}}_{B}$$
$$+ \underbrace{\sum_{\substack{u=1\\u\neq p}}^{p}\sum_{k=1}^{\beta} s_{i.u} x_{k.p} x_{k.u}}_{C} + \underbrace{\sum_{\substack{u=1\\u\neq p}}^{p}\sum_{k=1}^{\beta} s_{i.u} x_{k.u}\left(\frac{1}{N}\sum_{v=1}^{N} n_{vk}\right)^*}_{D}\}. \quad (8)$$

The expected value of the decision variable can be represented by:

$$E\left(D_{i.p}\right) = s_{i.p}\left(\frac{M{_}N}{M}\right) E_b \quad (9)$$

Since $n_{ik}$, ${}^{1}/_{N}\sum_{j=1}^{N} n_{vk}$ and $x_{k.p}$ are uncorrelated and independent, conditional variance of $D_{i.p}$ can be expressed as:

$$V\left(D_{i.p}\right) = E(A^2) + E(B^2) + E(C^2) + E(D^2) \quad (10)$$

where $E(\cdot)$ and $V(\cdot)$ are the mean and variance operators, respectively. The variance of $A$ is given by:

$$V(A) = \left(\frac{N+1}{N}\right)\left(\frac{M-N}{M}\right) E_b N_0/2 \quad (11)$$

Likewise, for the term B we have

$$V(B) = \beta N_0^2/4N \quad (12)$$

According to the independency conditions we have:

$$V(C) = \left(\frac{M-N}{M}\right) E_b(P-1) \quad (13)$$

$$V(D) = \left(\frac{N+1}{N}\right)\left(\frac{M-N}{M}\right) N_0(P-1) \quad (14)$$

Since the bit energy is considered as a deterministic variable, $D_{i.p}$ at the output of the correlator is a Gaussian random variable

[23]. Using equations (9) and (10), the BER of the $p^{th}$ user can be derived by means of the Gaussian approximation as follows:

$$BER_{i,p} = \frac{1}{2} erfc\left(\frac{E\left(D_{i.p}\middle|s_{i.p} = +1\right)}{\sqrt{2V\left(D_{i.p}\middle|s_{i.p} = +1\right)}}\right) \quad (15)$$

With replacing (11), (12), (13) and (14) in (15), the BER expression of the SA-OFDM-DCSK is obtained as:

$$BER_{i.p} = \frac{1}{2} erfc\left(\left[\left(\frac{N+1}{N}\right)\frac{PMN_0}{(M-N)E_b} + \frac{\beta M^2 N_0^2}{2N(M-N)^2 {E_b}^2} + \left(\frac{2(P-1)M}{(M-N)E_b}\right)\right]^{-\frac{1}{2}}\right) \quad (16)$$

### C. Sub-carrier Allocated MU-OFDM-DCSK (SA-OFDM-DCSK)

In this part, we employ the Lagrangian relaxation approach to tackle the BER minimization problem adopting (16) as the optimization criterion. Since the complementary error function $(erfc)$ is a monotonically decreasing function, we use its argument $\mathbb{E}_{Multi-user}(N)$ as the objective function.

$$\mathbb{E}_{Multi-user}(N) = \left(\frac{N+1}{N}\right)\frac{PMN_0}{(M-N)E_b} + \frac{\beta M^2 N_0^2}{2N(M-N)^2 {E_b}^2} + \left(\frac{2(P-1)M}{(M-N)E_b}\right) \quad (17)$$

So, the BER minimization problem can be represented as:

$$\min_{N} \mathbb{E}_{Multi-user}, \quad N \in \mathbb{Z}_+ \quad (18-\mathrm{a})$$

$$S.to: N < M. \quad (18-\mathrm{b})$$

Form the Lagrangian of problem with respect to the constraint we have:

$$\mathcal{L} = \left(\frac{N+1}{N}\right)\frac{PMN_0}{(M-N)E_b} + \frac{\beta M^2 N_0^2}{2N(M-N)^2 {E_b}^2} + \left(\frac{2(P-1)M}{(M-N)E_b}\right) + \mu[N-M] \quad (19)$$

We can form Karush-Kuhn-Tucker (KKT) conditions as follows [27]:

$$\sigma_{\mathcal{L}}/\sigma_N = 0 \quad (20-a)$$

$$\mu[N-M] = 0 \quad (20-b)$$

$$\mu \geq 0 \quad (20-c)$$

From $(20-b)$, $(20-c)$, we can conclude that $\mu = 0$. Thus, the problem in (18) is reduced to solving the following equation:

$$\sigma_{\mathbb{E}_{Multi-user}}/\sigma_N = 0 \quad (21)$$

By taking the first order derivation with respect to N and after some manipulations we have:

$$-2E_b(PN_0 + 2P - 2)N^3 + 2E_b(PMN_0 + 2PM - 2M - 2PN_0)N^2 + 3MN_0(2PE_b + \beta N_0)N - M^2 N_0(2PE_b + \beta N_0) = 0 \quad (22)$$

This is a cubic equation with the following general form:

$$AN^3 + BN^2 + CN + D = 0 \quad (23)$$

We use Cardano's method [23] to obtain the unknown variable $N$. To this aim, we first need to convert (23) into the following standard form in which the coefficient of $N^2$ vanishes. Thus, we embed $C = X - (B/3A)$ in (23).

$$X^3 + \zeta X + \xi = 0 \quad (24)$$

Then, we can write the desired form as [23]:

$$AX^3 + \left(\frac{-B^2}{3A} + C\right)X + \left(-\frac{B^3}{27A^2} - \frac{B}{3} + D\right) = 0 \quad (25)$$

Through dividing by A, the cubic equation in (25) can be represented as the desired form in (24) where the coefficients $\zeta$ , $\xi$ are given by:

$$\zeta = \frac{-(PM(N_0+2) - 2M - 2PN_0)^2}{3(PN_0 + 2P - 2)^2} + \frac{3MN_0(2PE_b + \beta N_0)}{2E_b(PN_0 + 2P - 2)} \quad (26)$$

$$\xi = \frac{(PM(N_0+2) - 2M - 2PN_0)^3}{27(PN_0 + 2P - 2)^3} + \frac{PM(N_0+2) - 2M - 2PN_0}{3(PN_0 + 2P - 2)} - \frac{(PM(N_0+2) - 2M - 2PN_0)}{2E_b(PN_0 + 2P - 2)} \quad (27)$$

We can form the discriminant $\Delta$ of the Cardano's formula, and obtain $X$:

$$\Delta = \frac{\zeta^3}{27} + \frac{\xi^2}{4} \quad (28)$$

$$X = \sqrt[3]{\frac{-\xi}{2} + \sqrt{\Delta}} + \sqrt[3]{\frac{-\xi}{2} - \sqrt{\Delta}} \quad (29)$$

Thus, $N^* = X - (B/3A)$ gives the solution of the original Equation (24) in $N^*$. In other words, the optimal number of reference sub-carriers for SA-OFDM-DCSK is the integer part of the obtained continuous variable $N$:

$$N^* = \left\lfloor \sqrt[3]{\frac{-\xi}{2} + \sqrt{\Delta}} + \sqrt[3]{\frac{-\xi}{2} - \sqrt{\Delta}} + \frac{PM(N_0+2) - 2M - 2PN_0}{3(PN_0 + 2P - 2)} \right\rfloor \quad (30)$$

## III. Joint Power and Sub-Carrier Allocated MU-OFDM-DCSK (PSA-OFDM-DCSK) System

By considering the power of the sub-carriers as an optimization variable, the problem formulation leads to a higher degree polynomial Equation. The basic difficulty of the previous formulation in (18) is the non-linearity of the objective function and it can be quite hard to get an optimum solution. In this section, we offer an optimal joint sub-carrier and power allocation policy using classical fractional programing methods. Before formulating the problem, we must explain the energy relations followed by calculating a BER expression for the power and sub-carrier allocated MU-OFDM-DCSK (PSA-OFDM-DCSK) system.

### *A. BER Derivation of the PSA-OFDM-DCSK*

The energy per bit equation of the PSA-OFDM-DCSK scheme can be obtained as:

$$E_b = \left(\frac{M}{M-N}\right)\frac{\sum_{m=1}^{M} c_m}{M}\sum_{k=1}^{\beta} x_{k,p}^2 = \left(\frac{\sum_{m=1}^{M} c_m}{M-N}\right)\sum_{k=1}^{\beta} x_{k,p}^2 \quad (31)$$

where $c_m$ is the power assigned to each of the $M$ sub-carriers. Now, we must derive BER expression by considering power coefficients $\sqrt{a}$, $\sqrt{b}$ for data-bearing and reference signals, respectively. At the receiver side, each user has a correlator for information detection by calculating the following decision variable $D_{i.p}$:

$$D_{i.p} = Re\{\sum_{k=1}^{\beta} Y\left(k.i.f_{s_{pi}}\right).R(k)^*\} \;. \quad (32)$$

where $R(k)$ is the averaged reference signal and $Y\left(k.i.f_{s_{pi}}\right)$ is the spread data which are given by:

$$Y\left(k.i.f_{s_{pi}}\right) = \sum_{p=1}^{P} \sqrt{a}\, s_{i.p} x_{k.p} + n_{ik} \;. \quad (33)$$

$$R(k) = \sqrt{b}\, x_{k.p} + {}^{1}\!/_{N}\sum_{v=1}^{N} n_{vk} \;. \quad (34)$$

Using the mean and variance of decision variable and Gaussian approximation in the manner described in the previous section, the BER expression of the PSA-OFDM-DCSK scheme can be written as shown in (35) at the bottom of the page[1]. After some manipulation we obtain:

$$BER_{i.p} = \frac{1}{2} erfc\left(\left[\left(\frac{b+aP}{ab}\right)\left(\frac{N+1}{N}\right)\frac{N_0\sum_{m=1}^{M} c_m}{(M-N)E_b} + \frac{\beta N_0^2\left(\sum_{m=1}^{M} c_m\right)^2}{2abN(M-N)^2 {E_b}^2} + \left(\frac{2(P-1)}{(M-N)E_b}\right)\right]^{-\frac{1}{2}}\right) \quad (36)$$

### *B. Joint power and sub-carrier allocation problem*

We consider the following BER minimization problem with a constraint $C_T$ on the total transmit power:

$$\min_{a,b,N} BER_{i.p} \quad (37-a)$$

$$S.to: \sum_{m=1}^{M} c_m \le C_T \quad (37-b)$$

$$a,b,N > 0. \quad (37-c)$$

Since the erfc (.) is a monotonically decreasing function, the problem (37) changes to a maximization one. We get $\sum_{m=1}^{M} c_m = (M-N)a + Nb$ for simplification. In the previous section, we started with (16) as the optimization criterion. This time we begin with (35) and accept $U$ in (38), as the objective function because this is a single rational function and can be handled by the classical form of fractional programing method [24]:

$$\max_{a,b,N} U(a,b,N) = \frac{\mathbb{A}(a,b,N)}{\mathbb{B}(a,b,N)} \quad (39-a)$$

$$S.to: (M-N)a + Nb \le C_T \quad (39-b)$$

$$a,b,N > 0, \quad (39-c)$$

where $\mathbb{A}(a,b,N)$ and $\mathbb{B}(a,b,N)$ are the nominator and denominator of the objective function, respectively. Among various algorithms for the fractional programing, there is an important class of algorithms which makes use of the following auxiliary parameter $q$:

$$\max_{(a,b,N)} V(a,b,N,q) = \{\mathbb{A}(a,b,N) - q\mathbb{B}(a,b,N)\} \quad (40)$$

$$S.t.: (39-b), (39-c).$$

An optimal solution of $V(a,b,N,q)$ is an optimal solution of $U(a,b,N)$. Thus, solving $U(a,b,N)$ is essentially equivalent to finding $q^*$ with $V(a,b,N,q^*) = 0$. For this purpose $V(a,b,N,q^*)$ has helpful properties such as continuity and convexity [25]. Thus, we form the Lagrangian function of the problem (40) as:

$$\mathcal{L}(a,b,N,q) = \mathbb{A}(a,b,N) - q\mathbb{B}(a,b,N) + \mu[(M-N)a + Nb - C_T] + \lambda a + \vartheta b + \theta N \,, \quad (41)$$

where $\mu$, $\lambda$, $\vartheta$, $\theta$ are Lagrangian multipliers. Using KKT conditions [27] we have:

$$\sigma_{\mathcal{L}}/\sigma_N = 0 \quad (42-a)$$

$$BER_{i.p} = \frac{1}{2} erfc\left(\sqrt{\frac{2Nab(M-N)^2 {E_b}^2}{(M-N)E_b\sum_{m=1}^{M} c_m\left[2(aP+b)(N+1)N_0 + +4ab(P-1)N\right] + \beta N_0^2\left(\sum_{m=1}^{M} c_m\right)^2}}\right) \quad (35)$$

$$U = \frac{2Nab(M-N)^2 {E_b}^2}{(M-N)E_b\sum_{m=1}^{M} c_m\left[2(aP+b)(N+1)N_0 + +4ab(P-1)N\right] + \beta N_0^2\left(\sum_{m=1}^{M} c_m\right)^2} \quad (38)$$

$$\sigma_{\mathcal{L}}/\sigma_b = 0 \quad (42-b)$$

$$\sigma_{\mathcal{L}}/\sigma_N = 0 \quad (42-c)$$

$$\mu[(M-N)a + Nb - C_T] = 0 \quad (42-d)$$

$$\lambda a = 0\,,\vartheta b = 0, \theta N = 0 \quad (42-e)$$

$$\lambda \geq 0,\mu \geq 0,\vartheta \geq 0,\theta \geq 0 \quad (42-f)$$

$$a,b,N > 0. \quad (42-g)$$

From (42-e) and (42-f) we have $\lambda = 0, \vartheta = 0, \theta = 0$. Assume that $(M-N)a + Nb < C_T$ and $\mu = 0$. From (42-a) and (42-b) and nulling the first derivative of the Lagrangian function with respect to $a$ , $b$, the optimal power of reference and data-bearing sub-carriers for each user is obtained in $q$. These equations are given in (43) and (44) at the bottom of the page[2]. In addition, nulling the derivative of the Lagrangian function with respect to $N$ results in the following quadratic equation:

$$A_1N^2 + A_2N + A_3 = 0, \quad (45)$$

where $A_1$, $A_2$ and $A_3$ are obtained as (46- 48) at the bottom of the page. From (43), (44), (45), and calculating of $q^*$ using bisection method, we can determine the optimal solutions $a^*$,$b^*$,$N^*$. It is known that [24], [26]:

$$V(a,b,N,q^*) > 0 \qquad \text{if } q < q^*, \quad (49-a)$$

$$V(a,b,N,q^*) = 0 \qquad \text{if } q = q^*, \quad (49-a)$$

$$V(a,b,N,q^*) < 0 \qquad \text{if } q > q^*. \quad (49-a)$$

Among various popular classical methods for solving nonlinear equations, we may note here the Newton method and the binary search method (Bisection method) [28-30]. The $q^*$ can be determined efficiently using Bisection search. To describe the bisection method, assume that an interval $[q^l\ q^h]$ containing $q^*$ is at hand. For this purpose, $q^l = 0$ and $q^h =$ (a large number) may be used, or $q^h$ may be computed. Corresponding to the outcome $V(a,b,N,q^*) > 0$ or $V(a,b,N,q^*) < 0$ for the middle point $q = 1/2\,(q^l + q^h)$, the interval $[q\ q^h]$ or $[q^l\ q]$ is respectively considered as the next interval to be tested. In summary, as indicated in Algorithm 1, the Bisection procedure can be described as follows:

Algorithm 1 : Compute $a^*$,$b^*$,$N^*$ by the Bisection method

Initialize a small number $\varepsilon$ as stop criterion.

**Step 1**: Find $q'$ and $q''$ such that $V(a,b,N,q') > 0$ and $V(a,b,N,q'') < 0$. Let $q^l$ : = q' and $q^h$ :=q ".

**Step 2**: Let $q = 1/2\,(q^l + q^h)$ , calculate a, b, and N. If $|V(a,b,N,q)| < \varepsilon$, halt. Otherwise go to Step 3.

**Step 3**: If $V\big(V(a,b,N,q)\big)$>0, let $q^l$:= q and return to Step 2. Otherwise let $q^h$: = q and return to Step 2.

## IV. Results and Discussions

In this section, we first present the numerical simulations to validate the effectiveness of the SA-OFDM-DCSK system, in which all users are assumed to have equal transmit power and the noise variance is assumed to be one. Then, we produce several numerical results for the performance evaluation of the PSA-OFDM-DCSK scheme.

### A. *Effectiveness of the SA-OFDM-DCSK system*

Fig. 4 shows the BER performance of an SA-OFDM-DCSK system with $M = 64$ , $\beta = 128$, and numerically simulated BER surface for the various number of references $N$ and the different number of users, $P$, under the equal power allocation assumption. We extract this plot regarding (16). Observe that for a single user scenario, when $N$ increases from 1, the BER decreases and reaches an optimal point $N^* = 12$, and increases again. As a result, the BER of the SA-OFDM-DCSK system can be minimized according to the variable $N$. The BER is a monotonically increasing function of the number of users, $P$.

In Fig. 5, the BER curves of the SA-OFDM-DCSK is plotted for different $E_b/N_0$ values using the analytical sub-carrier allocation expression in (16). There is a good agreement between theoretical predictions in Section 3. For example, when $N$ increases from 1 to 3, the BER performance can drop to about $10^{-2}$ at the target $E_b/N_0 = 9\ (dB)$ due to reference diversity

$$a = \frac{NbE_b{}^2(M-N)^2 - qM(N+1)(M-N)E_bN_0b - q\beta N_0^2NbM - 2qN^2b^2E_b(P-1)(M-N)}{2q(N+1)(M-N)^2E_bN_0P + q\beta N_0^2(M-N)^2 + 4q(M-N)^2NbE_b(P-1)} \quad (43)$$

$$b = \frac{(M-N)a\big[N(M-N)E_b{}^2 + q(N+1)E_bN_0(N-PN-M) - q\beta N_0^2N - 2qNaE_b(P-1)(M-N)\big]}{q\beta N_0^2N^2 + qN[(N+1)(M-N)E_bN_0 + 4NaE_b(P-1)(M-N)]} \quad (44)$$

$$A_1 = 3\big\{abE_b{}^2 + q(a-b)[(ap+b)E_bN_0 - 2ab(P-1)E_b]\big\} \quad (46)$$

$$A_2 = \big\{-2MabE_b{}^2 + 2q(aP+b)E_bN_0(2Ma - Mb + a + b) - q\beta N_0^2(a+b)^2 + 4abE_b(P-1)qM(a-b)\big\} \quad (47)$$

$$A_3 = M\big\{abE1 +_b{}^2\ M - q(aP+b)E_bN_0(Ma - 2b - 2b) + q\beta N_0^2a(a+b) - 2a^2bE_b(P-1)qM\big\} \quad (48)$$

and when P increases from 2 to 3, the BER performance is degraded because of the multi-user interference effects.

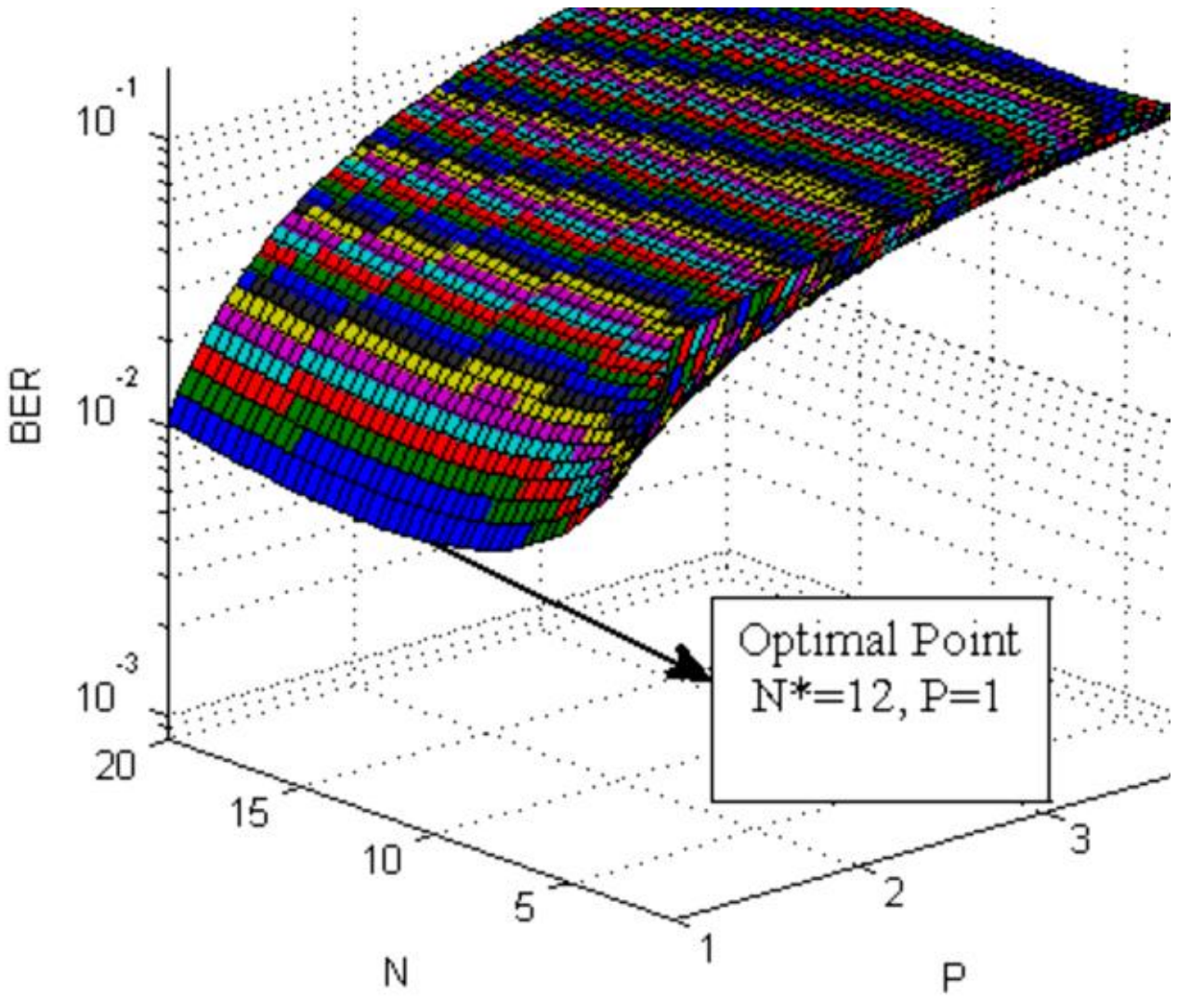

Fig. 4. The BER surface of SA-OFDM-DCSK system versus N, P with $E_b/N_0 = 10\ dB$.

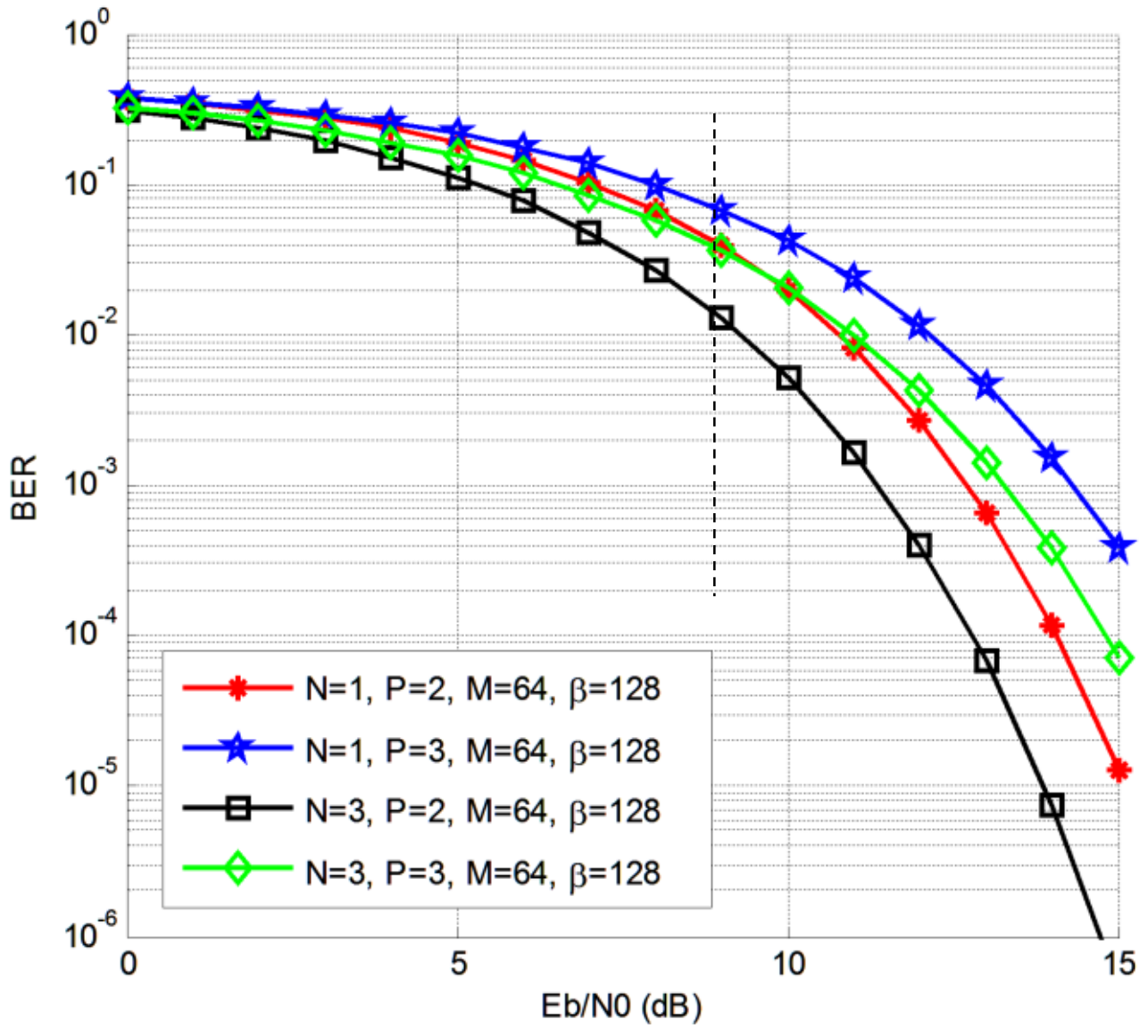

Fig. 5. Analytical BER performance for different $E_b/N_0$ for various $P$, $N$.

We compare the analytical performance of the SA-OFDM-DCSK system with conventional MU-OFDM-DCSK in [11]. As emphasized in Fig. 6, the BER performance of the sub-carrier allocated SA-OFDM-DCSK system is superior to the regular MU-OFDM-DCSK cases with the different non-optimal $N$. It implies that the reference number is a decisive parameter and we can obtain a non-coherent OFDM-DCSK system with a better BER performance by considering the optimal number of reference sub-carriers. In other words, the superiority of curve with $N^* = 12$, validates the optimality of the achieved closed form solution. For instance, when N increases from 1 to 12, the BER gain is about 0.1 at the target $E_b/N_0 = 11\ (dB)$.

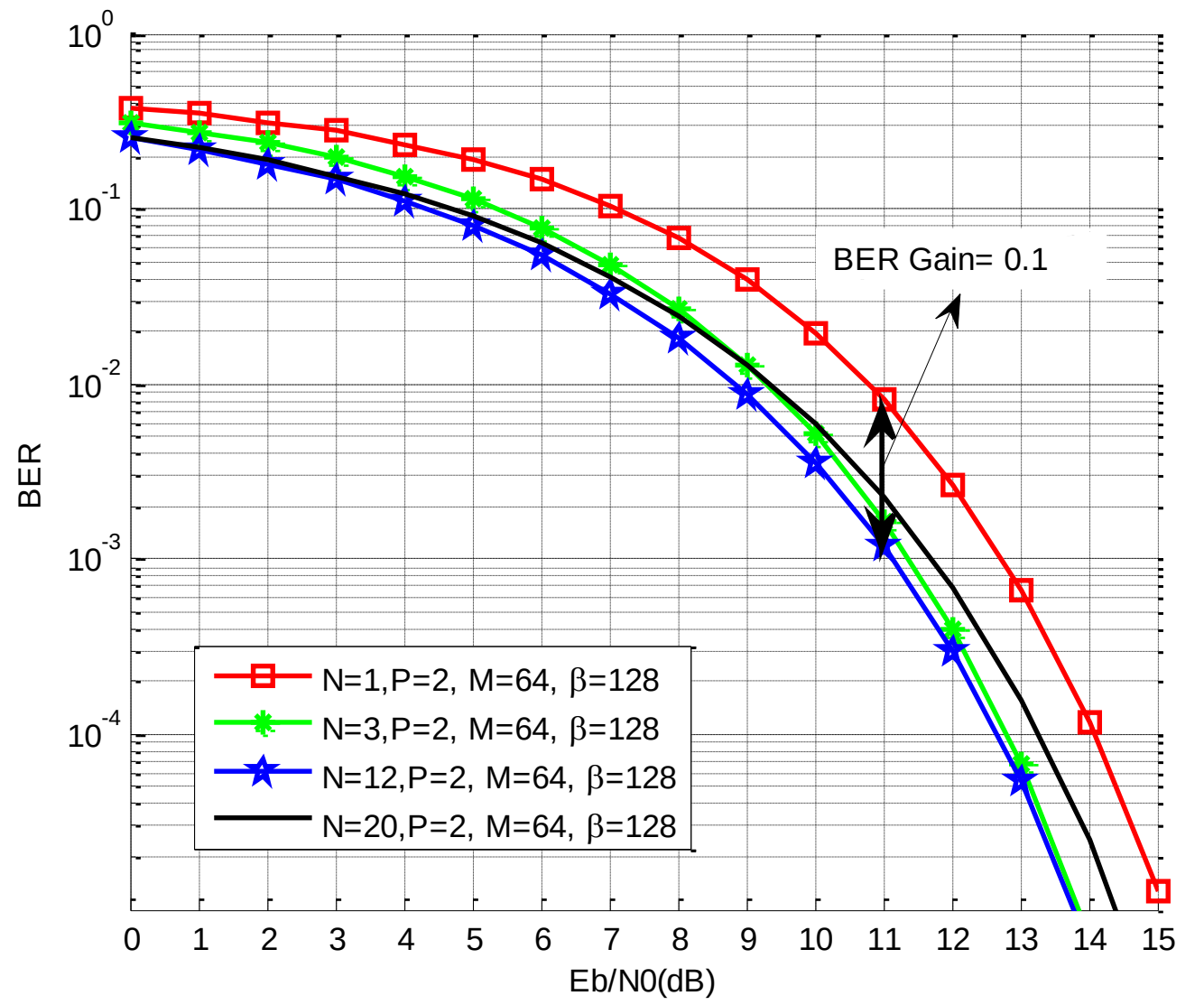

Fig. 6. Performance comparison of the optimal system with the non-optimal one, for different $E_b/N_0$.

## B. *Effectiveness of the PSA-OFDM-DCSK system*

In Fig. 7, the BER performances of the PSA-OFDM-DCSK system is evaluated using (35) with $M = 64$ , $\beta = 128$. The BER surface is plotted for the different number of $N$, $a$, and b. The unique optimum point located at the bottom of the surface indicates that the result obtained by the algorithm converges to a minimum point. If we compare the performance of this system with the SA-OFDM-DCSK system presented in Figure 4, because the power allocation is performed and more power is given to the references, the number of required optimal references ($N^* = 4$) is much less than the SA-OFDM-DCSK ($N^* = 12$), and this leads to better spectral efficiency.

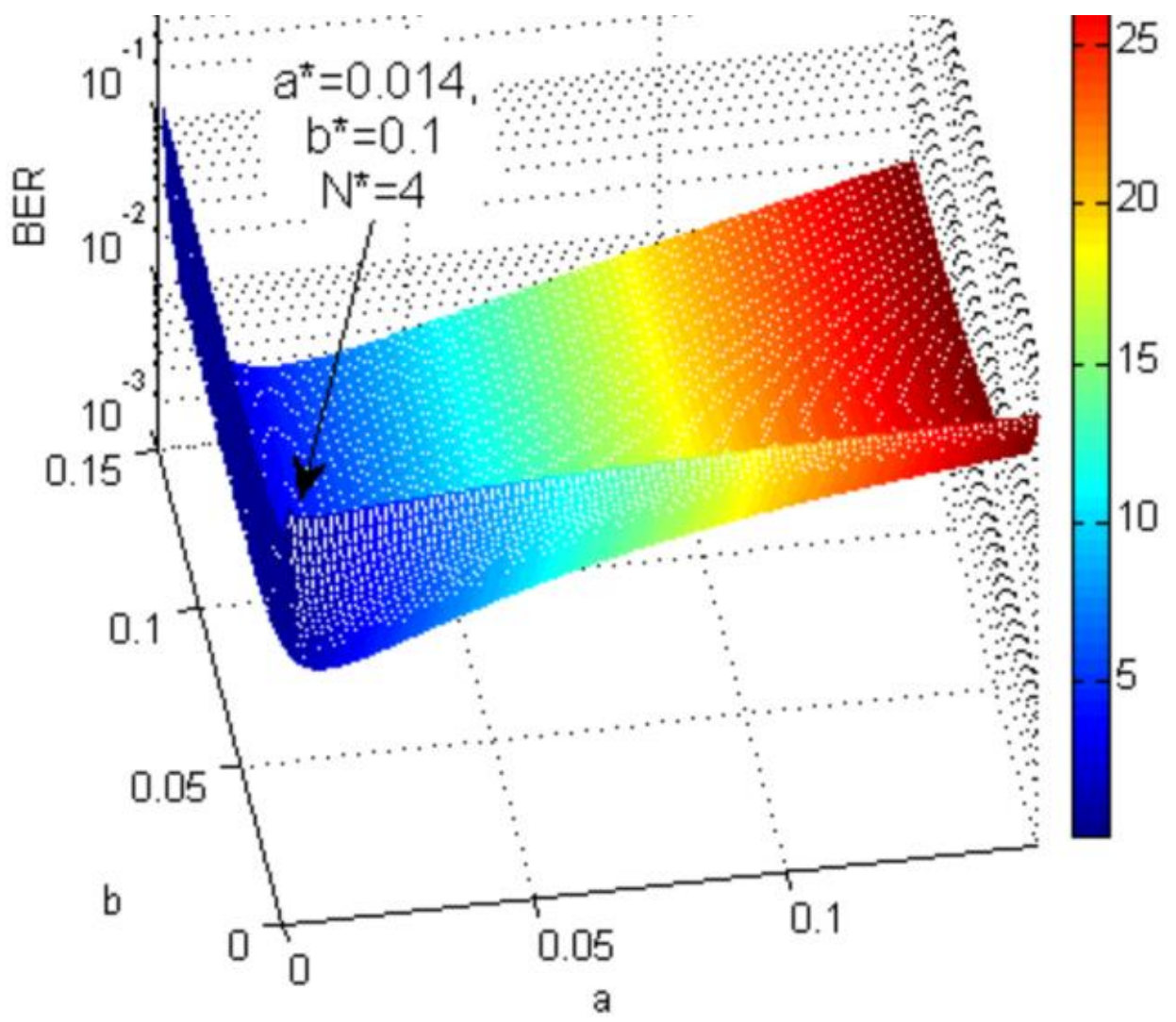

Fig7. BER surface of PSA-OFDM-DCSK system versus N, a , b, with $E_b/N_0 = 10\ dB$.

Fig.8 is a direct comparison between the PSA-OFDM-DCSK approach and SA-OFDM-DCSK systems with $M = 64$ , $\beta = 128$, for the different number of users. Observe that the achieved BER using the PSA-OFDM-DCSK approach is significantly superior to the SA-OFDM-DCSK with equal power allocation. For instance, when the number of users is $P =$ 3, we have a BER gain about 0.1 at the target $E_b/N_0 = 10\ (dB)$. Furthermore, when the number of users increases from $P = 2$ to $P = 3$, the performance reduction of the PSA-OFDM-DCSK is less than the case we use the SA-OFDM-DCSK system.

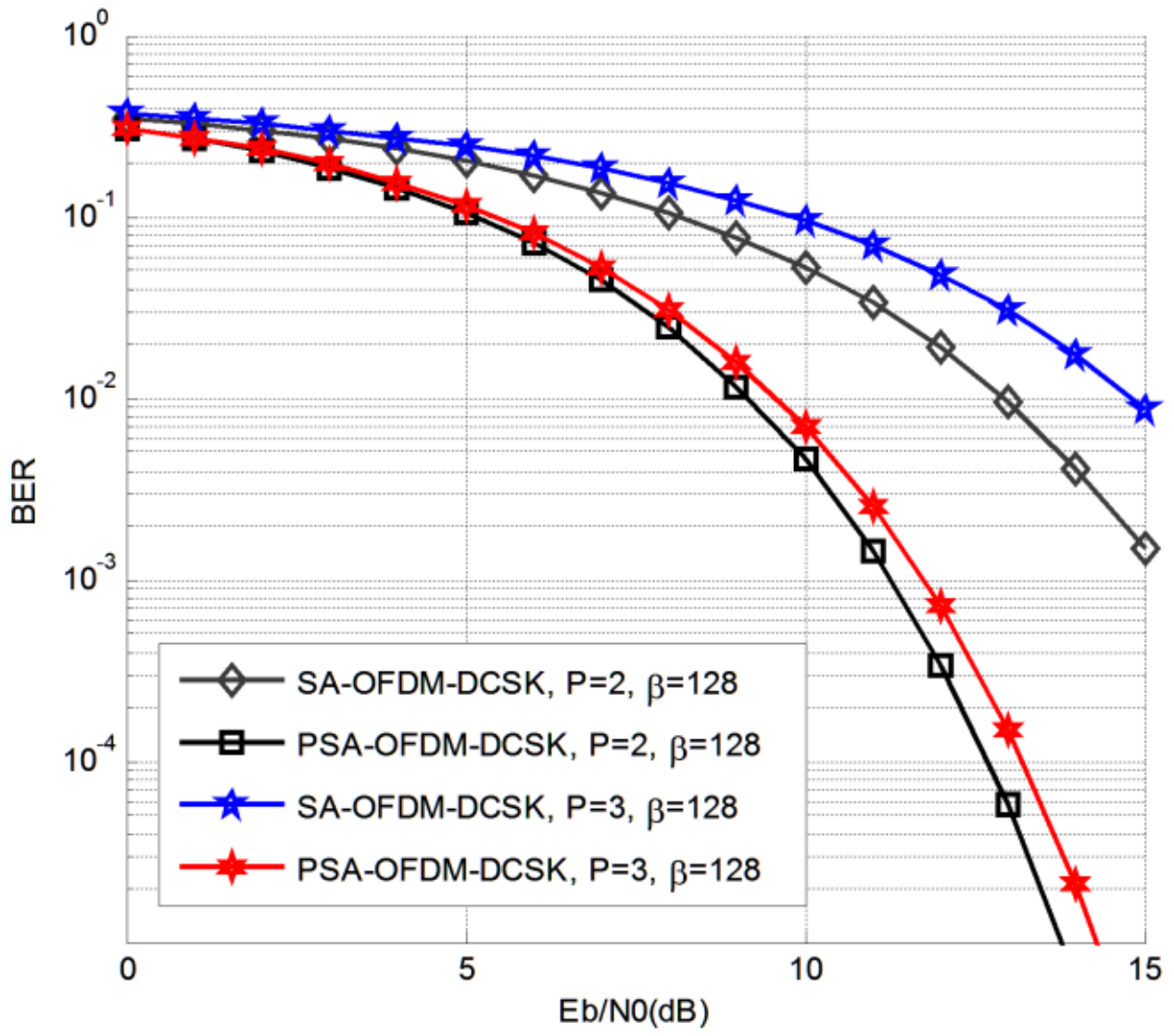

Fig8. The effect of $P$ on the performance of the SA-OFDM-DCSK and PSA-OFDM-DCSK systems.

Fig. 9 compares the theoretical BER curves of the two schemes for the spreading factors 16, 64,128, when $M = 64$ and $P = 2$. An increase in $\beta$ results in a little degradation in BER performance for the SA-OFDM-DCSK scheme and this degradation is obviously bigger for the PSA-OFDM-DCSK.

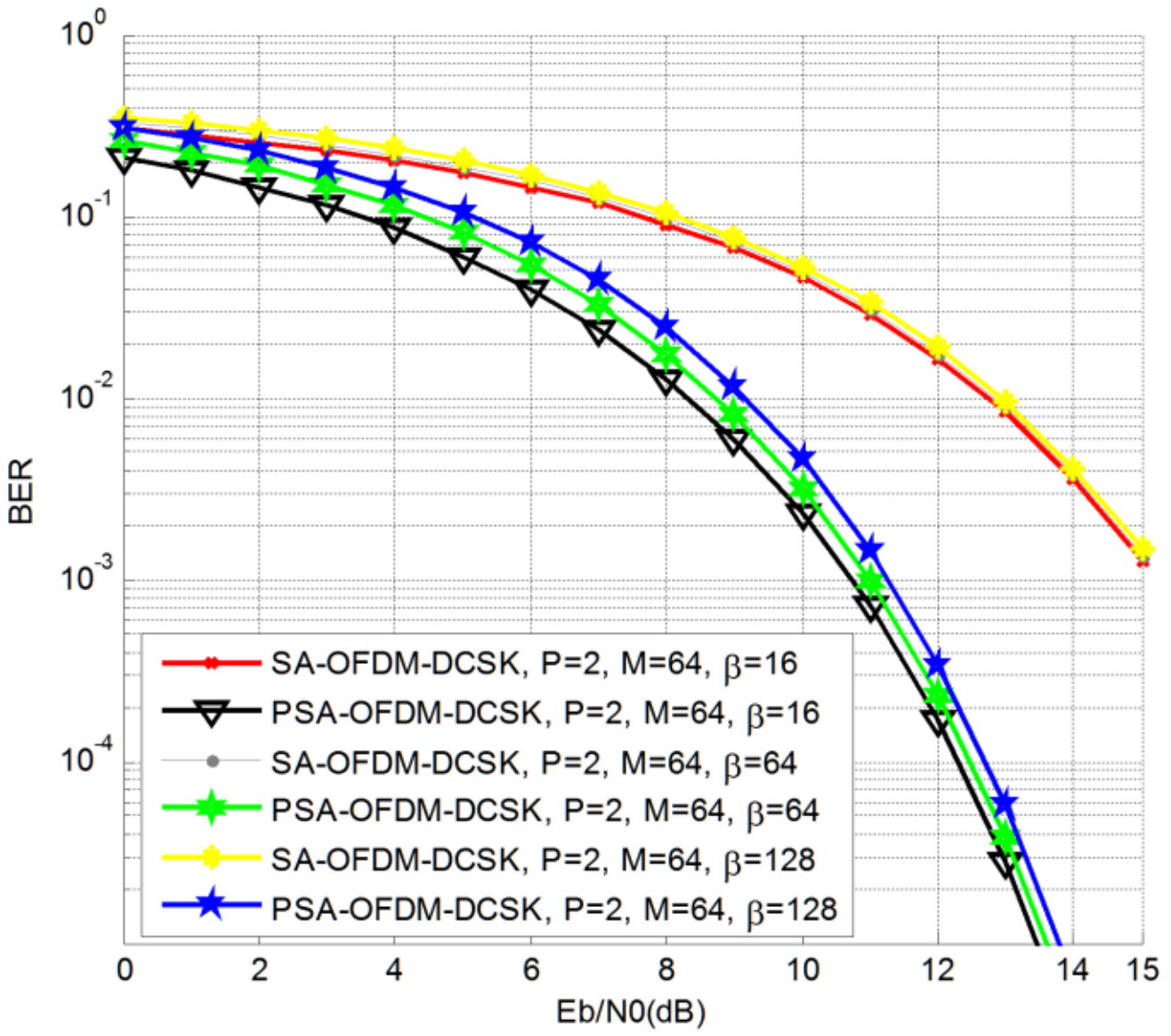

Fig. 9 Analytical BER performances of the SA-OFDM-DCSK and PSA-OFDM-DCSK systems for various $\beta$.

Fig. 10, indicates the effects of $N$ on the performance of the PSA-OFDM-DCSK for $\beta = 128$, M=64, $E_b/N_0 = 10\ (dB)$, and $P = 2$, under a fixed reference power $b = 0.01$. For $N = 6$, the BER performance can only reach about $0.5 \times 10^{-2}$. If we choose the optimal number of sub-carriers, $N = 3$, the BER performance drops to about $2 \times 10^{-3}$ at the target power $a = 0.01$.

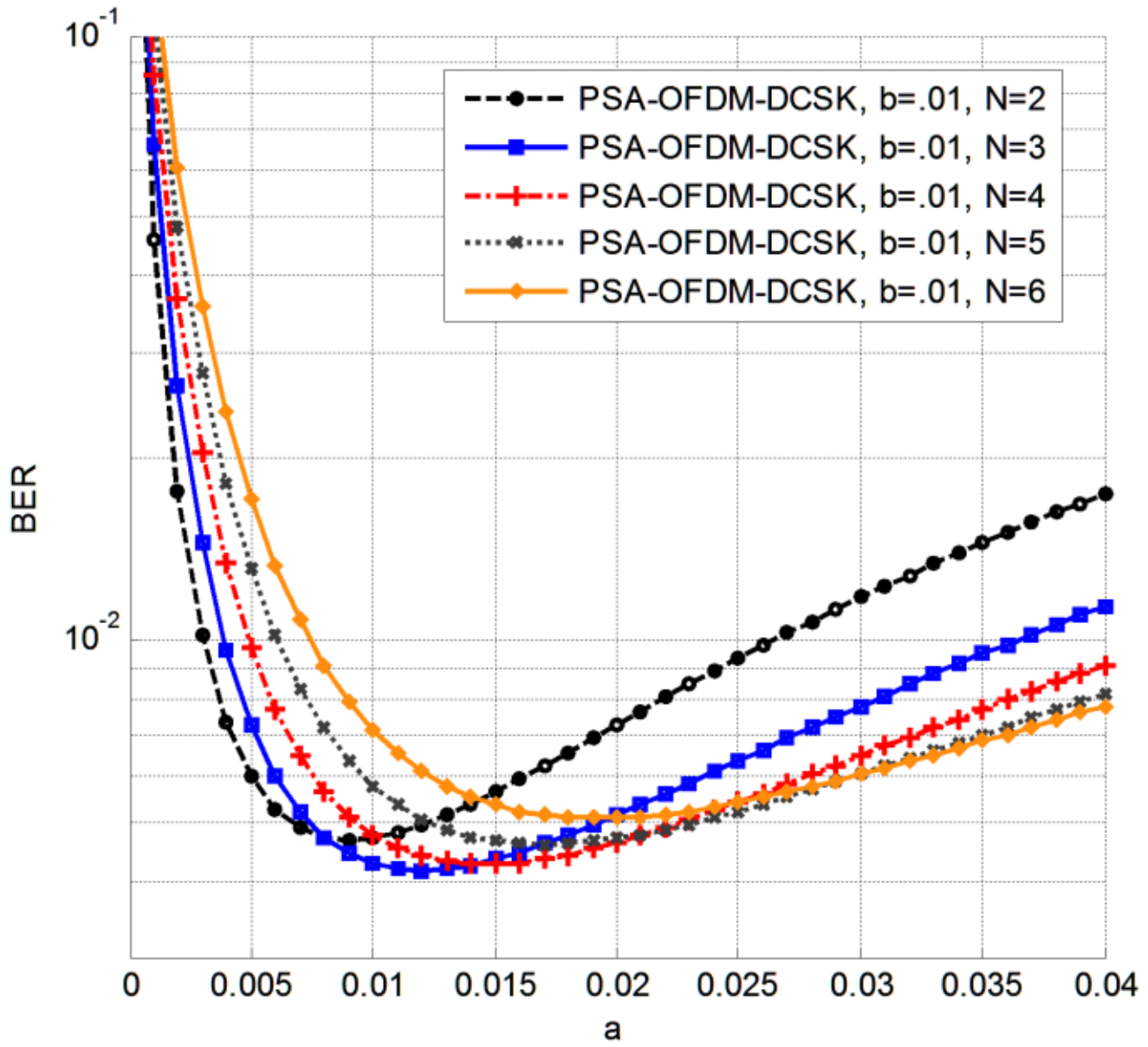

Fig. 10. The effects of $N$ on the performance of PSA-OFDM-DCSK for $\beta = 128$, M=64, $E_b/N_0 = 10\ (dB)$, $P = 2$.

In Fig. 11, the effectiveness of the power allocated to the data-bearing sub-carriers (variable $a$) in the PSA-OFDM-DCSK system is investigated for a fixed number of sub-carriers ($N^* = 3$) under the AWGN channel. Observe that the optimal power is obtained $a^* = 0.005$, for $\beta = 128$, $M = 64$, $P = 2$ and given $E_b/N_0 = 10\ (dB)$.

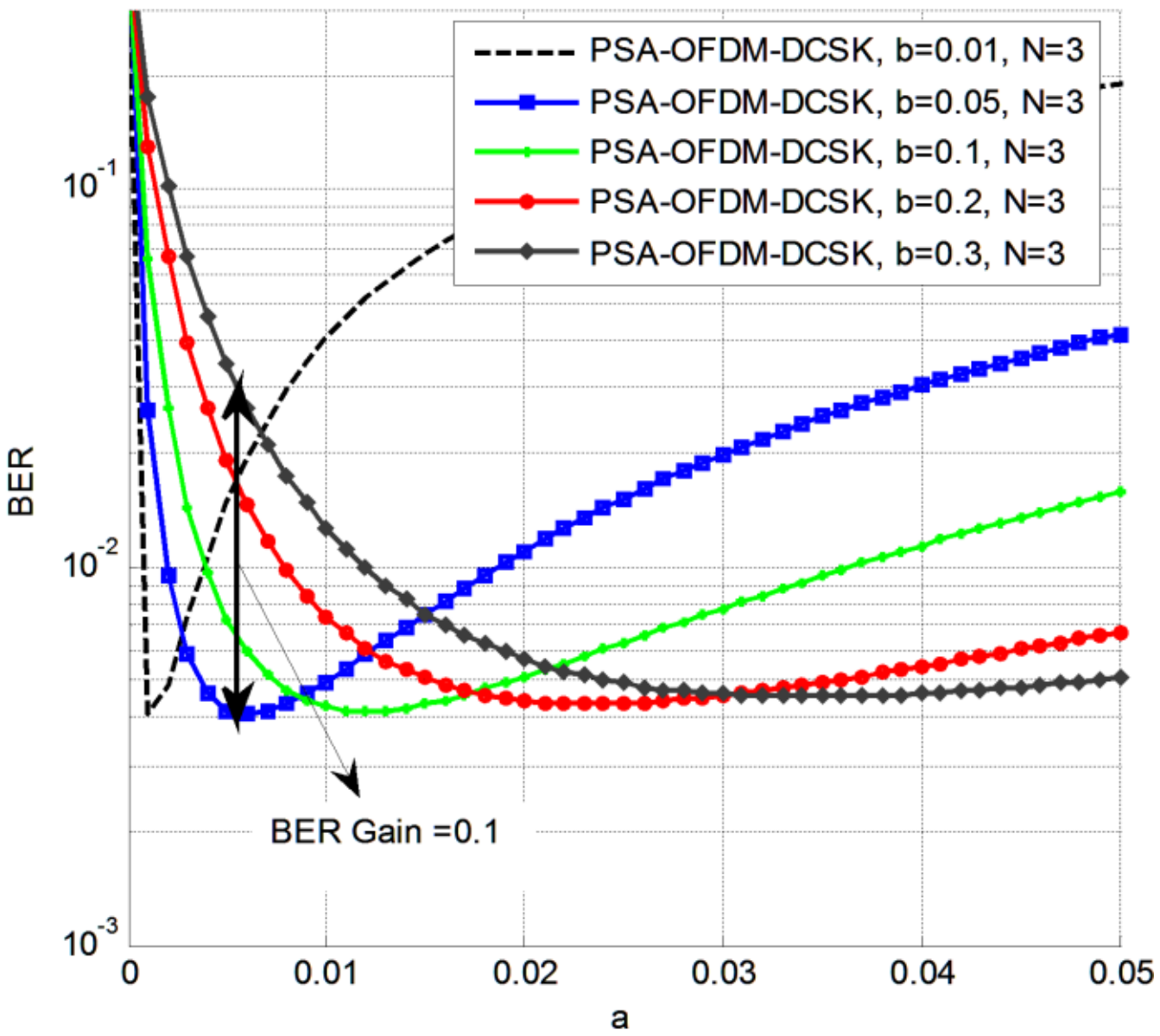

Fig. 11. The effects of optimal power allocation on the performance of the PSA-OFDM-DCSK system for $\beta = 128$, M=64, $E_b/N_0 = 10\ (dB)$, $P = 2$.

The BER curve has an inflection with respect to $a$. Hence, the optimal power allocation strategy obtains better BER performance for smaller $a$. When the power of reference subcarriers decreases from $b = 0.3$ to $b = 0.05$, the gain of BER is about 0.1 at the target $a^* = 0.005$. By taking into consideration the optimal power allocation, transmitters might not always require to operate at their largest power and the ratio between the power of the references and data-bearing sub-carriers is an important determiner.

## V. Conclusions

In this study, we have investigated the resource allocation strategies for MU-OFDM-DCSK systems. Particularly, we have presented two resource allocation policies to i) Minimizing the BER performance of each user by determining the optimal number of the reference sub-carriers. ii) Jointly sub-carrier and power allocation by formulating the BER minimization as a fractional problem. We used Lagrangian relaxation for the first problem and found the optimal number of sub-carrier as a closed-form solution using Cardano's method. We also have proposed a jointly sub-carrier and power allocation approach (PSA-OFDM-DCSK) by conversion of a fractional objective function into a subtractive form. We provided an algorithm based on the bisection approach to determine most appropriate number of references and the optimal power assigned to them. Simulation results suggest that the PSA-OFDM-DCSK outperforms both the SA-OFDM-DCSK and conventional MU-OFDM-DCSK systems in terms of the BER performance. Determining the maximum number of users and testing power allocation strategies for multi-user scenarios seems to be interesting research fields for future studies.

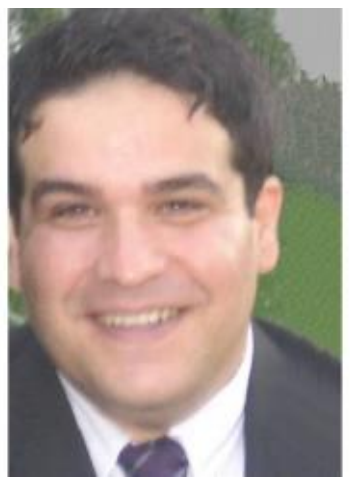

**M. Mobini**, received his M.Sc. degree in communications engineering with the first rank from the Department of Electrical and Computer Engineering, Amirkabir University of Technology (Tehran Polytechnic), Tehran, Iran, in 2013, and the Ph.D. degree in communications engineering from the Babol Noshirvani University of Tech- nology, in 2019. He is currently a Research Associate with the Department of Electrical and Computer Engineering, Babol Noshirvani University of Technology. His current research interests include wireless Communication, optimization, deep learning, Internet of Things, cyber-physical systems, biomedical instruments, and interdisciplinary researches.

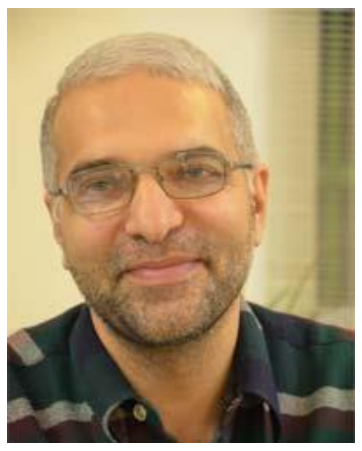

**M. R. Zahabi,** received his Bachelor and Post graduate degrees in 1990 and 1993 respectively in Electrical Engineering from K.N TOOSI University of Technology and Amir Kabir University of Technology. He received his Ph.D. degree in 2008 in electrical Engineering from Université de Limoges, France. He has several journal and conference publications in national and international level. Currently, he is a faculty member at Babol University of Technology and he has guided a large no. of graduates. His current areas of interest are wireless communication and MIMO, Coding, Analog Decoders, UWB systems and RFID.